\documentclass[twoside]{article} 
\usepackage{fleqn,espcrc2} 
\usepackage{epsfig}

\newcommand{\gsim}{\;\raisebox{-0.9ex}
                   {$\textstyle\stackrel{\textstyle>}{\sim}$} \;}
\def\bp {{ \mathbf{p} }} 
\def\ls {{\ell s}} 
\def\os {{os}} 
\def\ot {{\otimes}}
\def\lg {\left\langle}
\def\rg {\right\rangle}

\title{
\begin{flushright} 
\small HEPHY--PUB 732/2000 \\ 
\small STPHY 46/00
\end{flushright}
Multiplicity dependence of Bose-Einstein correlations 
        in $\bar{p}p$ reactions: \\
        a discussion of possible origins\thanks{
Proceedings of the 9th International Workshop on 
Multiparticle Production, Torino, June 11-18, 2000,
edited by  A.\ Giovannini and R.\ Ugoccioni;
Nuclear Physics B Supplement (to be published).} 
}  

\author{B.\ Buschbeck\address{Institut f\"ur Hochenergiephysik,
                       Nikolsdorfergasse 18, A--1050 Vienna, Austria}
        and H.C.\ Eggers\address{Department of Physics, University of
                       Stellenbosch,7600 Stellenbosch, South Africa}
}

\begin{document}

\begin{abstract}
  The observed pronounced multiplicity dependence of correlation
  functions in hadron-hadron reactions and in particular of
  Bose-Einstein correlations provides information about underlying
  physics. We discuss in this contribution several interpretations,
  giving special attention to the string model for Bose-Einstein
  correlations of Andersson and Hofmann, as well as the core-halo
  picture of Cs\"org\H o, L\"orstad and Zimanyi.
\end{abstract}

\maketitle

\setcounter{footnote}{0}

\section{Introduction}

It has been well known for years \cite{alt,UA1,TEV} that Bose-Einstein
(BE) correlations measured in hadron-hadron reactions at sufficiently
high energies exhibit a pronounced dependence on multiplicity in the
form of the strength parameter $\lambda$. In $e^+e^-$ reactions, this
behaviour is either absent \cite{Aleph,Lowe} or at least much weaker,
especially with a 2-jet selection \cite{Op}. Heavy ion reactions, on
the other hand, see $\lambda$ values decreasing with multiplicity at
lower multiplicity densities (lower $A$ reactions) but increasing
again\footnote{
  It should be borne in mind, though, that $\lambda$ measurements of
  different experiments are not easily compared because of varying
  systematics such as the chosen fit function and fit range, the
  ``cleanness'' of the pion sample, different treatments of Coulomb
  corrections etc. The described tendencies in $\lambda$ are therefore
  more qualitative than quantitative in nature.}
at higher multiplicity densities \cite{Carlos}.

These findings may suggest a possible common explanation in terms of
superposition of several sources or strings, where each of them
symmetrizes separately according to the model of Andersson and
Hofmann\cite{And86a}. No BE correlations are then expected between
decay products of different strings \cite{And97a}. The following
arguments could be used: In the framework of the Dual Parton Model for
hadron hadron and heavy ion reactions \cite{DPM}, it is expected that
higher multiplicities correspond to a higher number of strings or
chains. This can explain the multiplicity dependence in hadron-hadron
reactions. In $e^+e^-$ reactions, with a selection of two-jet events,
only one single string is formed and consequently one does not expect
much dependence of BE correlations on multiplicity. In heavy ion
reactions, there is again an increasing number of strings with
increasing multiplicity density, but eventually the densely-packed
strings coalesce until they finally form a large single fireball. This
picture can qualitatively explain why the BE signal is decreasing at
low $A$ and finally increasing again \cite{Carlos}.  A clean situation
of a superposition of two strings occurs if two $W$-bosons are
coproduced in $e^+e^-$ reactions and both decay hadronically. If there
are no BE correlations between the pions from different strings, the
$\lambda$ values are expected to be diminished.  Unfortunately, these
measurements are hampered by low statistics and experimental
difficulties \cite{WWD,WWA,WWL3}.

It should be stressed, however, that this is not the only possible
explanation of experimental behaviour. In the case of hadron-hadron
reactions, for example, several alternative explanations do exist
\cite{MD2000}, among them the core-halo picture \cite{CSO} which
connects consistently the multiplicity dependence of correlation
functions of like-charge with those of opposite-charge pion pairs.

From all these considerations, it should be clear that there is
important information in the observed multiplicity dependencies, in
particular when comparing different reaction types.

This contribution concentrates on a discussion of possible origins
which could lead in the case of high energy hadron-hadron reactions to
the observed multiplicity dependence. We first adress in Section 2 the
question of the influence of jet production by using again $\bar{p}p$
collisions at $\sqrt{s} = 630$~GeV measured in the UA1 detector
\cite{MD2000,UA1_89a}.  Low-$p_T$ and high-$p_T$ subsamples are
investigated.  Using the low-$p_T$ subsample where the influence of
jets is removed to a large extent, we finally discuss in Section 3
possible underlying physics.

\section{The influence of jets}

It is well known \cite{UA1-jet} that in hadron-hadron reactions the
probability of jet production rises with energy and multiplicity. The
multiplicity dependence of correlation functions can be influenced by
this hard subprocess. To investigate this influence, we used a data
sample similar to that in \cite{MD2000} but with larger statistics: It
consists of 2,460,000 non-single-diffractive $\bar{p}p$ reactions at
$\sqrt{s} = 630$~GeV measured by the UA1 central
detector~\cite{UA1_89a}. Only vertex-associated charged tracks with
transverse momentum $p_T \geq 0.15$~GeV/c and $|\eta | \leq 3$ have
been used. We restrict the azimuthal angle to $45^\circ \leq |\phi |
\leq 135^\circ$ (``good azimuth'').  These cuts define a region in
momentum space which we call $\Omega$.

Since the multiplicity for the entire azimuthal range is the
physically relevant quantity, we select events according to their
uncorrected all-azimuth charged-particle multiplicity $N$. The
corrected multiplicity density is then estimated as twice $n$, the
charged-particle multiplicity in good azimuth: $(dN_c/d\eta) \simeq
2(dn/d\eta)$.

The measured quantities are $K_2^I$, the second order normalized
cumulant correlation functions in several multiplicity intervals $N
\in [A, B]$; for a complete definition, refer to \cite{MD2000}. They
are measured for pairs of like-sign ($\ls$) and opposite-sign ($\os$)
charge separately as
\begin{eqnarray}
\label{se}
\overline{K_2}^{I\;\ls}(Q| AB) \!\!\!
&=&  \!\!\!
{\overline{n}_{\ls}^2 \over \overline{{n(n{-}1)}}_{\ls}}
\; r_2^{\ls}(Q|AB) \; - \;  1 \,,
\\
\label{se1}
\overline{K_2}^{I\;\os}(Q |AB) \!\!\!
&=& \!\!\!
{\overline{n}_+ \, \,  \overline{n}_- \over \overline{n_+ n}_-}
\; r_2^{\os}(Q|AB) \; - \;  1 \,,
\end{eqnarray}
where $\overline{n}_{\ls}\ (= \overline{n}_+ = \overline{n}_-)$ ,
$\overline{n_+ n}_-$ and $\overline{{n(n{-}1)}}_{\ls}$ are the mean
numbers of positive or negative particles, $\os$ pairs and $\ls$ pairs
respectively in the whole interval $\Omega$ and in the multiplicity
range $[A,B]$.  The prefactors in front of the normalized density
correlation functions $r$ correct for the bias introduced by fixing
multiplicity \cite{MD2000}.  The functions $r_2^{\ls}$ and $r_2^{\os}$
are defined for $\ls$ and $\os$ pairs in the correlation integral
description \cite{coI}
\begin{eqnarray}
r_2^\ls (Q)
&=&  {\rho_2^\ls(Q) \over \rho_1\ot\rho_1^\ls (Q) }
     \\
&=&  {\rho_2^{++} (Q)
      \over
      \rho_1^+ \ot \rho_1^+ (Q)
     }
=    {\rho_2^{--} (Q)
      \over
      \rho_1^- \ot \rho_1^- (Q)
     }
     \,, \nonumber\\
r_2^{os} (Q)
&=& {\rho_2^\os(Q) \over \rho_1\ot\rho_1^\os (Q) }
    \\
&=& {\rho_2^{+-} (Q)
     \over
     \rho_1^+ \ot \rho_1^- (Q)
    }
=   {\rho_2^{-+} (Q)
     \over
     \rho_1^- \ot \rho_1^+ (Q)
    } \,, \nonumber
\end{eqnarray}
with $Q = \sqrt{-(p_1 - p_2)^2}$ the spacelike four-momentum
difference of the pion pairs.

Internal cumulants (\ref{se}) and (\ref{se1}) are analysed in
three samples as a function of pion transverse momentum $p_T$
as follows:
\begin{itemize}
\item[(i)] An all-$p_T$ sample of all like-sign and opposite-sign pion
  pairs in $\Omega$.  This is shown in Fig.\ 1 for three
  representative multiplicity densities.
  
\item[(ii)] A low-$p_T$ subsample containing only charged particles\footnote{
    The pion sample contains about 15\% kaons.}
  with $p_T \leq 0.7$~GeV/c. Also removed from the subsample were
  entire events containing either a jet with $E_{T} \geq 5$~GeV or at
  least one charged particle with $p_T \geq 2.5$~GeV/c.  This is shown
  in Fig.\ 2 with the same three multiplicity selections.  The
  prefactors entering Eqs.\ (\ref{se}),\ (\ref{se1}) are calculated in
  this case by using only charged particles $0.15 \leq p_T \leq
  0.7$~GeV/c.
  
\item[(iii)] A high-$p_T$ subsample where only charged particles with
  $p_T \geq 0.7$~GeV/c are considered. This is shown in Fig.\ 3. It
  has been demonstrated previously \cite{Clau} that all high-$p_T$
  particles stem from jets or minijets.
\end{itemize}

From Figs.\ 1--3, we see that whereas the low-$p_T$ subsample behaves
very similarly to the all-$p_T$ sample, there is a pronounced increase
in the strength of correlation functions for the high-$p_T$ case (iii)
(note the different scale on the plot!). This can be interpreted as
the influence of jets, which are inherently spiky in nature. In this
case, BE correlations are hence mixed up with correlations originating
from jets, so that it would be difficult to measure them separately.

We note further that $\ls$ functions in Fig.\ 3 reveal a crossover in
the region $Q \simeq 1$~GeV, making the determination of the
multiplicity dependence of $K_2^I$ highly $Q$-dependent. The fit to
this high-$p_T$ sample using eq.\ (\ref{ex1}) gives ``radius
parameters'' $R$ which decrease with increasing multiplicity, in
contrast to the samples (i) \cite{MD2000} and (ii).  Fig.~3 shows also
a pronounced secondary peak after a minimum at $Q \simeq 1.4$~GeV
which can be attributed to the onset of local $p_T$ compensation with
two back-to-back particles with at least $p_T = 0.7$~GeV/c
corresponding to the cut applied in this subsample \cite{csopriv}.

The fits performed to $\ls$ pair data in Fig.1 -- Fig.3 are
exponential,
\begin{equation}\label{ex1}
K_2^{I\, \ls} (Q)
= a + \lambda \ e^{-RQ}\,.
\end{equation}
The corresponding dependence of $\lambda$ on multiplicity is used in
subsequent figures.

Fig.\ 4 compares the multiplicity dependence of the all-$p_T$ and
low-$p_T$ samples in the small-$Q$ region ($Q = 0.1$~GeV). The two
samples differ only slightly in their respective cumulants as well as
their $\lambda$ values.

Fig.\ 5 shows the corresponding multiplicity dependence in the
high-$p_T$ sample, once again on a larger scale in $K_2^I$. The
influence of jets shows up dramatically: all correlation functions are
increased in height, and the $\os$ functions do not show a
multiplicity dependence for particle densities $dN_c/d\eta \leq 3.3$.
A decrease like in (i) and (ii) is probably compensated by increasing
jet activity.

The interplay of BE correlations and resonance production with the
onset of jet production and the transition from soft to hard
interactions are interesting questions in their own right and can be
studied with samples like the high-$p_T$ one. We will, however,
concentrate in the following on the low-$p_T$ subsample.

\section{Multiplicity dependence of the low-$p_T$ sample}

The results of a study with the all-$p_T$ sample (i) have been
published in Ref.~\cite{MD2000}. In Fig.~6, we plot for the low-$p_T$
sample (ii) the same ratio
\begin{equation}
{K_2^I\left(Q\; |\; (dN_c/d\eta = 6.9) \right)
 \over
 K_2^I\left(Q\; |\; (dN_c/d\eta = 1.2) \right) }
\end{equation}
for like-sign and opposite-sign pairs respectively, while Fig.~7 shows
the behaviour of the low-$p_T$ cumulants for fixed $Q$ but varying
multiplicity. These figures show that the all-$p_T$ results found
previously remain valid for the low-$p_T$ sample also, namely:
\begin{itemize}
  
\item[--] The like-sign and opposite-sign cumulants have very similar
  multiplicity dependence when compared in the same $Q$ region.
  
\item[--] Cumulants behave distinctly differently at small and large
  $Q$: at small $Q$, the multiplicity dependence of both samples is
  weaker than $1/N_c$, while at large $Q\ (\gsim 2$~GeV) the cumulants
  are negative and follow roughly a $1/N_c$ law\footnote{
    For simplicity we write $N_c$ instead of $dN_c/d\eta$ here and
    below ($N_c = 6 \cdot dN_c/d\eta$).}.
  
\item[--] A third region around $Q = 1$~GeV shows small and rapidly
  changing cumulants.
\end{itemize}
In the following, we discuss four possible explanations of these
phenomena. It should be stressed that the LUND Monte Carlo model
(PYTHIA) cannot reproduce the multiplicity dependence of correlation
functions and in particular of the BE effect \cite{Fang}.
\begin{itemize} 
\item[3.1] {\bf Bose Einstein correlations are the result of
    symmetrization within individual strings only} \cite{And86a}: When
  several strings are produced, each string symmetrizes separately and
  decay products of different strings would hence not contribute to BE
  correlations \cite{And97a}.
  
  Because unnormalized cumulants of independent distributions combine
  additively, the independent superposition in momentum space of $\nu$
  equal sources/strings, each with a $q$th order cumulant
  $\kappa_q(\bp_1,\ldots,\bp_q)$ and each with some multiplicity
  distribution (e.g.\ Poisson) results in an unnormalized combined
  cumulant $\kappa_q^{(\nu)}$ of
\begin{equation}
\kappa_q^{(\nu)}(\bp_1,\ldots,\bp_q)
= \nu \; \kappa_q(\bp_1,\ldots,\bp_q) \,,
\end{equation}
while the combined single particle spectrum is given in terms of
individual sources' spectra by
\begin{equation}
\rho_1^{(\nu)}(\bp)  
= \nu \; \rho_1(\bp) \,,
\end{equation}
so that the $q$th order normalized cumulants for $\nu$ superimposed
sources is given by
\begin{eqnarray}
\label{ss1}
K_q^{(\nu)}(\bp_1,\ldots,\bp_q)
\!\!\!\!&=&\!\!\!\! {\nu \; \kappa_q(\bp_1,\ldots,\bp_q)  
   \over 
   \nu^q \; \rho_1(\bp_1)\ldots,\rho_1(\bp_q)}
 \\
&=&\!\!\!\!
 {1 \over \nu^{q-1}} \; K_q(\bp_1,\ldots,\bp_q) \,. \nonumber
\end{eqnarray}
Hence $K_2^{(\nu)}$ is inversely proportional to the number of
sources.  This remains true for the correlation integral
$K_2^{(\nu)}(Q)$ also.

The above derivation is only for illustration. If $K_2^{(\nu)} \propto
1/\nu$, this would imply $K_2^{(\nu)} \propto 1/N$ only if $N \propto
\nu$, i.e.\ for identical sources each of fixed multiplicity. In
reality, the assumption of equal sources is probably not fulfilled.
The following scenario might be more realistic: ``Fixing multiplicity
does not necessarily mean fixing the number of sources. The sources
(we will them define below) probably do possess a whole multiplicity
distribution rather than a single fixed multiplicity.  Our selected
multiplicities range from 0.83 to 9.1, varying over about a factor 10.
At small $Q$, however, the $K_2^{\ls}$ vary only by at most a factor
3. This suggests that at the highest selected multiplicity we would
observe the superposition of only 3 sources, from which we are
sampling their high multiplicity tails.''
\footnote{This, however, implies no multiplicity dependence of
  correlation functions within one source, which still is a strong
  assumption.}
In the Dual Parton Model approach \cite{DPM}, one source might be
identified with the topology of one pomeron exchange.\footnote{
  In the DPM, the exchange of one pomeron corresponds already to the
  formation of two chains or strings.  We consider this case here as
  ``one source''.}
If we select low-multiplicity events, we expect to select the case of
one pomeron exchange.  Multiparton collisions corresponding to
multipomeron exchange are expected to contribute to
higher-multiplicity events.  Estimates in ref. \cite{Ma99} predict
two- to three-pomeron exchanges at the highest multiplicities seen by
UA1. The number of sources would increase correspondingly.  This could
explain the suppression of $\ls$ (Bose-Einstein) functions in Fig.~7a.
However, additional assumptions are needed to explain the similar
behaviour of $\ls$ and $\os$ functions in Figs.~6a and 7a and their
$1/N$ dependence at large Q ( Figs 6b, 7b).  Resonance production and
colour reconnection effects might be candidates (see sect. 3.3).

\item[3.2] {\bf Quantum statistical approach}: A chaoticity parameter
  $p$ decreasing with multiplicity would, in the quantum statistical
  approach \cite {marky,And93a}, decrease BE correlations at higher
  multiplicities.  In this picture, however, the question arises what
  the physical nature of the subprocess causing increased coherence at
  higher multiplicities would be. Also, the similarity of the
  behaviour of $\ls$ and $\os$ correlation functions in Fig.~7 is not
  easily explained within this framework.

\item[3.3] {\bf BE and resonance-induced correlations combined}: One
  could hypothesise that the observed multiplicity dependence of $\ls$
  correlation functions is the result of two processes \cite{Bi00}:
  
  a) Bose-Einstein correlations in the classical sense which, being a
  global effect, are independent of multiplicity \cite{wei}
  \footnote{The observed multiplicity dependence in ref. \cite{wei} is
    compensated by our prefactors in Eqs.\ (\ref{se}) and \ 
    (\ref{se1}).}, and
  
  b) the production of higher mass-resonances or clusters decaying
  into two or more like-sign pions: $R^* \to \pi^{\pm} \pi^{\pm} + X$
  (as seen for example in $\eta^{\prime}$ decay). If the unnormalized
  cumulants $\kappa_2^{I\, \os}$ and $\kappa_2^{I\, \ls}$ were wholly
  the result of resonance decays and if the number of resonances were
  proportional to the multiplicity $N_c$, then $\kappa_2^I \propto
  N_c$.  Assuming $\rho_1(\bp\,|N_c) \propto N_c\, \rho_1(\bp)$ gives
  $\rho_1 \ot \rho_1 \propto N_c^2$, and hence after normalization,
  the resonance-inspired guess yields $1/N_c$ behaviour,
  \begin{equation} \label{res1}
  K_2^I  
  = {\kappa_2^{I\, {\rm res}} \over \rho_1\ot\rho_1}
  \propto {1 \over N_c} \,.
  \end{equation}
  A mixture of processes a) and b) would give
  \begin{equation} \label{resin}
  K_2^{I\, \ls} (Q\,|N_c)
  \approx a(Q) + {b(Q) \over N_c} \,,
  \end{equation}
  in agreement with the behaviour of $\ls$ functions in Fig.~7a
  (straight line). The $1/N_c$ dependence in Fig.~7b at large $Q$
  suggests the existence of a resonance/cluster component for both
  $\ls$ and $\os$ pair production. The resonance contribution to the
  correlation functions would be concentrated mainly at the region
  around $Q\simeq 1$~GeV as in the case of $\rho^{0}$ production,
  which is visible as a peak in the $\os$ functions in Figs.\ 1--3.
  This region is however difficult to investigate because there the
  $K_2^{I\,os}$ are decreasing rapidly with increasing $Q$ while the
  $K_2^{I\,\ls}$ are already small. A $1/N_c$-dependence due to
  resonances or clusters in this dominant phase space region around 1
  GeV could presumably cause the large-$Q$ region to follow suit via
  missing pairs, thus explaining the observed dependence there.

  Once again, however, the similarity of $\ls$ and $\os$ functions in
  the small-$Q$ region in Fig.\ 7a can hardly be explained by assuming
  only the two components a) and b).

  One possible explanation could be the existence of global
  correlations for os pairs too.  It would give a behaviour for
  $K_2^{I\,\os}$ similar to that of Eq.~(\ref{resin}) and would
  explain its constant $a(Q)$ part by noting that the number of
  ${+}{-}$ pairs that can be formed from $N_+$ positive pions and
  $N_-$ negative pions would be $N_+ N_-$. If each of these pairs was
  correlated (statistically speaking), i.e.\ if all $(N_+ + N_-)$
  pions were mutually correlated, the unnormalised $\kappa_2^{I\,\os}$
  would be proportional to $N_c^2$ and hence after normalisation
  $K_2^{I\,\os}$ constant in $N_c$. Such a constant term signalizes
  maximum possible correlations in some events.

  More generally and following the arguments leading to
  Eq.~(\ref{res1}), we could also say: ``If resonance production were
  to rise more quickly than $\propto N_c$, then $K_2^{Ios}$ would
  decrease more slowly than $1/N_c$''.

\item[3.4] {\bf The Core-Halo picture} \cite{CSO}.  This picture is
  currently the only one which connects the multiplicity dependence of
  $\ls$ with that of $\os$ correlation functions.  The core-halo
  picture is based on the fact that Bose-Einstein correlations of
  decay products from long-lived resonances are not observable by
  experiments because they occur below experimental resolution and
  hence by definition belong to the ``halo'' of resonances that decay
  at large distances.  Examples are $1/ \Gamma_{\omega} = 23.5$~fm/c,
  $1/\Gamma_{\eta^\prime} = 986.5$~fm/c, $1/\Gamma_{\eta} =
  164400$~fm/c, which are all unresolvable within the UA1 experiment
  which can resolve decay products for distances $\leq 6$~fm only,
  corresponding to $Q \geq 30$~MeV.  Because of the halo, the
  $\lambda$ values of BE fits are in general reduced \cite{CSO},
  \begin{equation} \label{rla}
\lambda(Q_{min}) = f_c^2 \; {P^2_{1,{\rm core}}(p) \over P^2_1(p)} \,,
\end{equation}
where the second factor describes the momentum dependence and
$Q_{min}$ is the smallest $Q$ value accessible by the 
experiment\footnote{
In \cite{CSO} is assumed that $Q_{min} \simeq 10$~MeV and
  that the fit parameters (of Gaussian fits) are insensitive to the
  exact value of $Q_{min}$ in a certain restricted region.}.
The relation between $\lambda$ and the fraction of pions emitted by
the halo follows from Eq.\ (\ref{rla}) and from
\begin{equation} \label{rla1}
f_c = {\lg N_{\rm core}\rg  \over \lg N\rg } 
= 1- { \lg N_{\rm halo}\rg  \over \lg N\rg }
\,,
\end{equation}
where $\lg N_{\rm core}\rg$ is the mean multiplicity of directly
produced ``core'' pions, $\lg N_{\rm halo}\rg $ is the mean
multiplicity of halo pions and $\lg N\rg $ = $\lg N_{\rm core}\rg $ +
$\lg N_{\rm halo}\rg $. This means that
\begin{equation} \label{rla2}
\lambda \propto {\lg N_{\rm core}\rg^2  \over \lg N\rg^2 }
= \left( 1- { \lg N_{\rm halo}\rg  \over \lg N\rg } \right)^2
\,.
\end{equation}
From Eq.\ (\ref{rla2}) it is evident that $\lambda$ remains
independent of multiplicity only if $\lg N_{\rm core}\rg \propto \lg
N\rg $ and consequently $\lg N_{\rm halo}\rg \propto \lg N\rg $.  If
however the number of halo-resonances increases faster than $ \propto
\lg N\rg$, then also the number of their decay particles $\lg N_{\rm
  halo}\rg$.  As a consequence the BE parameter $\lambda$ will
decrease with multiplicity.

A second consequence emerges immediately (see last sentence in sect.
3.3): The fraction of $K_2^{Ios}$ stemming directly from the decay of
halo resonances will decrease less rapidly than $\propto 1/N$.  A
previous study \cite{pod} revealed the $Q$ region where two-body
$\pi^+\pi^-$ decay products of the halo resonances $\eta$,
$\eta^\prime$ and $\omega$ contribute, namely in $0.03 < Q \leq
0.55$~GeV. This is exactly the region where $\os$ correlation
functions indeed show a multiplicity dependence weaker than $\propto
N^{-1}$ as shown in Fig.~6b (where the $1/N$ case is indicated by the
dashed line).

So far, this discussion is purely qualitative. How the decrease of
$\lambda$ with $N_c$ would compare to an effective slower-than-$1/N_c$
decrease of $K_2^{I\,\os}$ is, of course, a quantitative question not
answered by the above argument, both because $\lambda$ refers to the
minimum $Q$~-value and its stability against shifts is not yet tested,
and because the experimental fraction $\lg N_{\rm halo}\rg / \lg N\rg$
is known only sparsely, if at all.  A more quantitative estimate has
to be done in future, including the fact that the previously measured
$K_2^{\ls}$ for the whole sample \cite{Eg97} are already near unity at
$Q \simeq 0.03$~GeV, even after subtracting the background
contribution due to the non-Poissonian overall multiplicity
distribution.  Correcting for an additional halo contribution could
finally cause $K_2^{\ls}$ for ${Q \to 0}$ to be greater than 1.
\end{itemize}

The four different explanations considered above each have some merit.
It is clearly desirable to shorten this list of candidates. We believe
that higher order cumulants are suitable for this purpose: If
symmetrization of individual strings is the right explanation, we can
expect from Eq.\ (\ref{ss1}) and Ref.\ \cite{GA} that the higher-order
cumulants would decrease much faster with multiplicity than $K_2^I$
(e.g.\ $K_3^I \propto 1/\nu^2$ ). Since such a fast decrease is not
predicted e.g.\ for the core-halo picture \cite{csopriv}, the
measurement of the multiplicity dependence of $K_3^I$ could probably
decide between the two cases (3.1) and (3.4).

\section*{Acknowledgements} 

\noindent
We thank B.\ Andersson, A.\ Bia\l as, T.\ Cs\"org\H o, 
K.\ Fia\l kowski and W.\ Kittel for useful discussions, and thank also the UA1
collaboration for freely providing the data.  We gratefully
acknowledge the technical support of G.\ Walzel.  HCE thanks the
Institute for High Energy Physics in Vienna for kind hospitality. This
work was funded in part by the South African National Research
Foundation.



\begin{figure*}[t]
  \centerline{\epsfig{figure=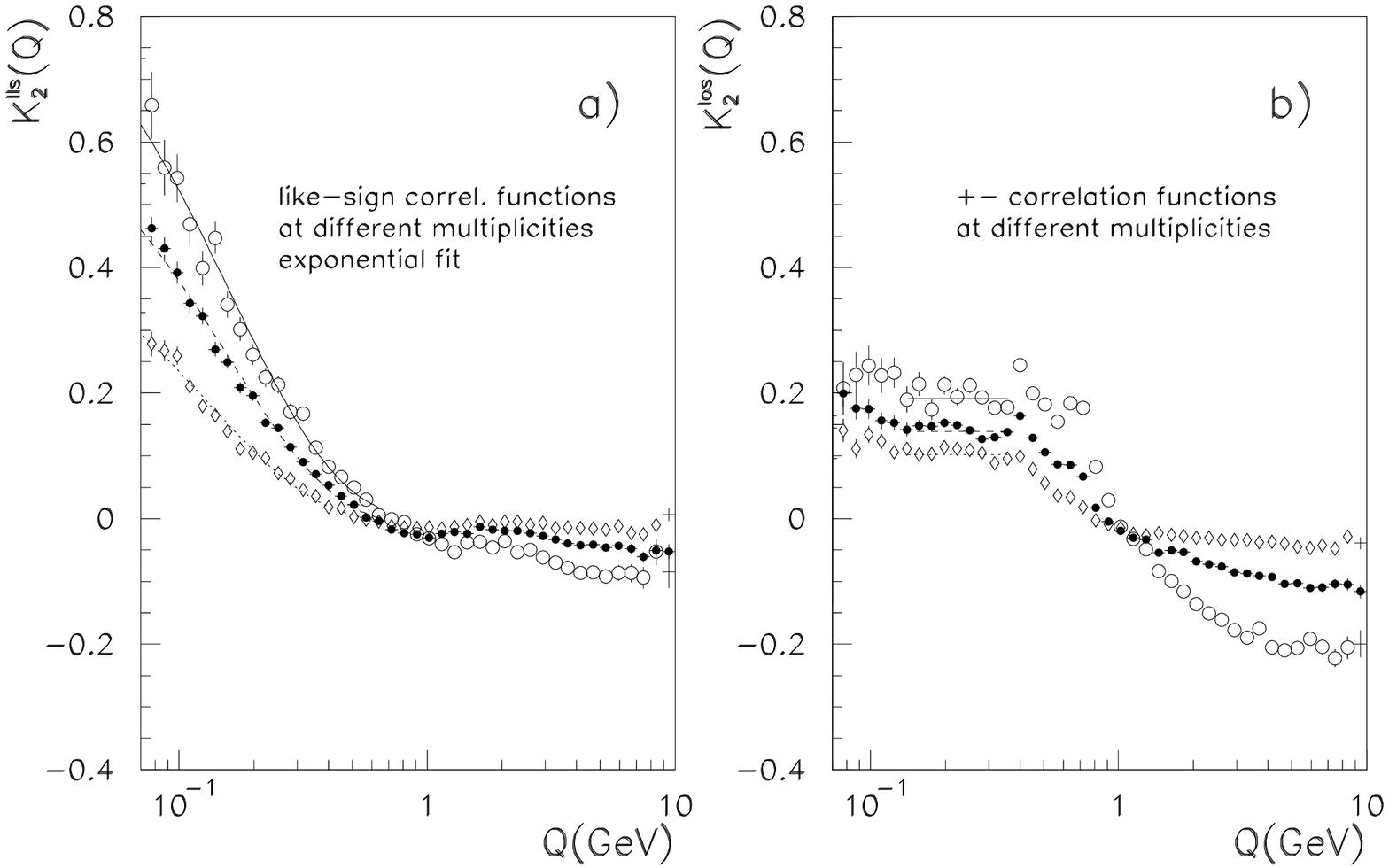,width=140mm}}
  \caption{
    Normalised $\ls$ and $\os$ internal cumulants for the all-$p_T$
    sample (i) with multiplicity densities $dN_c/d\eta = 1.22$ (open
    circles), $dN_c/d\eta = 2.72$, (full circles) and $dN_c/d\eta =
    6.85$ (diamonds).  The fits in a) are exponential as in eq.\ 
    (\ref{ex1}).  }
\end{figure*}

\begin{figure*} 
  \centerline{\epsfig{figure=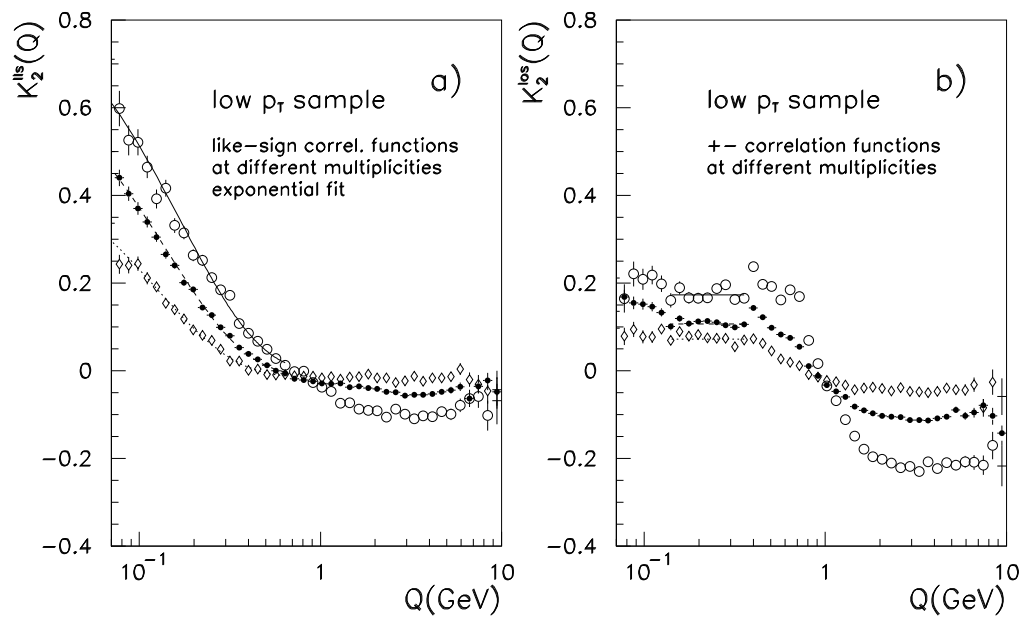,width=140mm}}
  \caption{
    Same as in Fig.\ 1, but for the low-$p_T$ sample (ii).  }
\end{figure*}

\begin{figure*}
  \centerline{\epsfig{figure=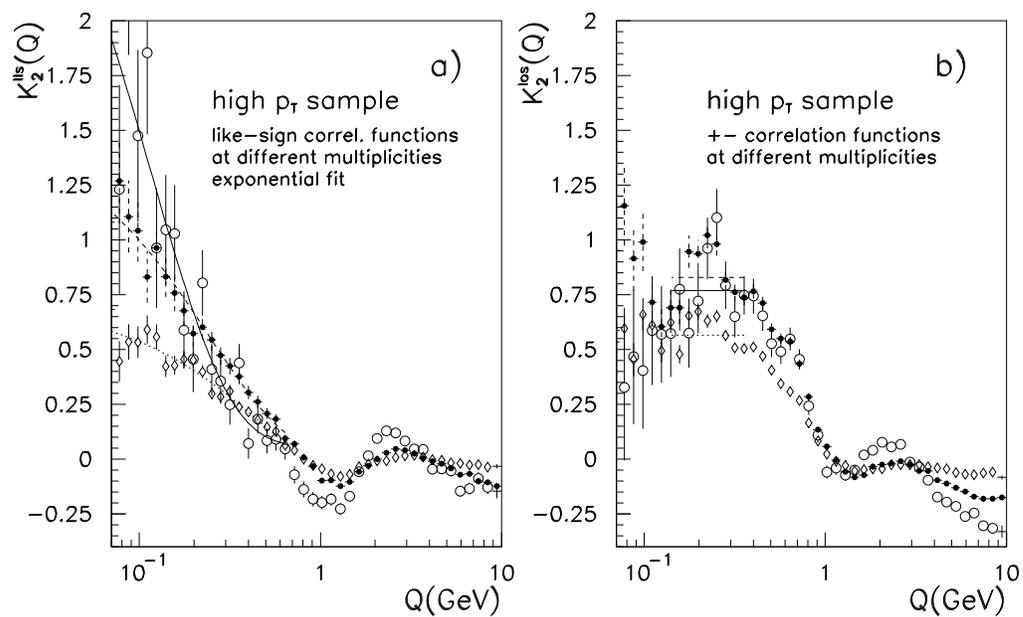,width=140mm}}
  \caption{Same as Fig.1, but for the high-$p_T$ sample (iii).}
\end{figure*}

\begin{figure*}[t]
  \centerline{\epsfig{figure=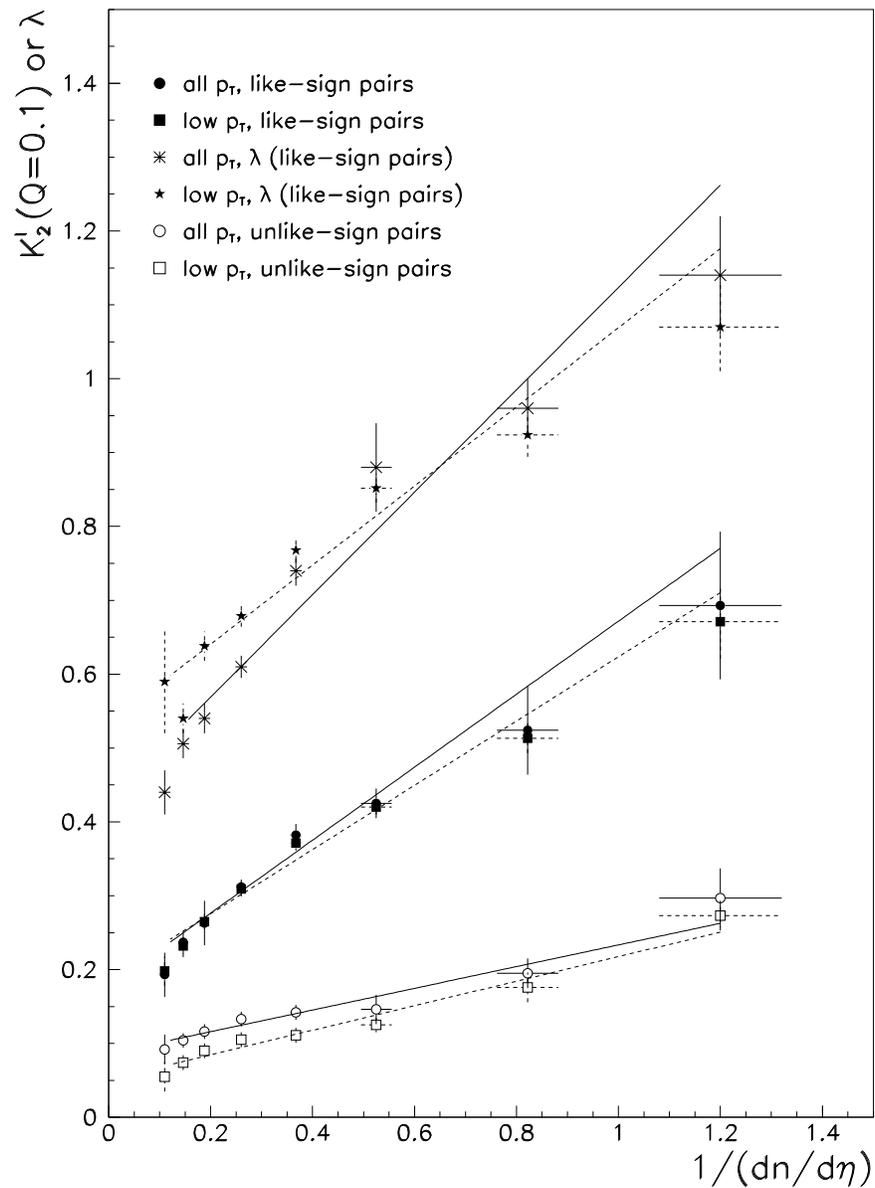,width=122mm}}
  \caption{
    Comparison of the multiplicity dependence of the low-$p_T$ sample
    (ii) with that of the all-$p_T$ sample (i). Full lines are fits of
    the all-$p_T$ sample to the $1/N_c$ behaviour as in eq.\ 
    (\ref{resin}); dashed lines are the corresponding fits for the
    low-$p_T$ sample.}
\end{figure*}

\begin{figure*}
  \centerline{\epsfig{figure=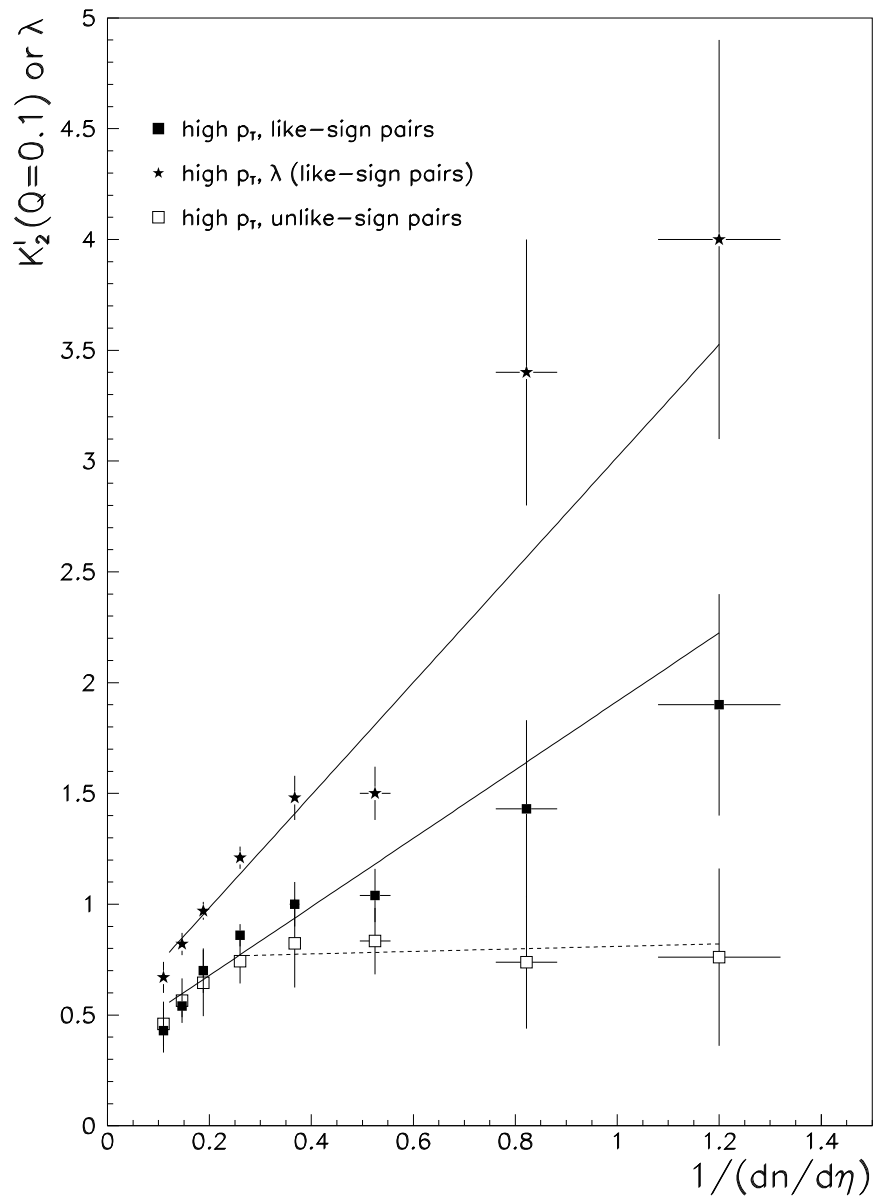,width=122mm}}
  \caption{
    Multiplicity dependence of the high-$p_T$ sample (iii).  The lines
    (full lines for $\ls$ pairs, dashed line for $\os$ pairs and for
    $1/(dN_c/d\eta) > 0.2$), pro-forma fits using (Eq.\ 
    (\ref{resin})), are clearly an inadequate representation of the
    data and are hence intended only to provide comparison to $1/N_c$
    behaviour. Note that these high-$p_T$ cumulants exceed one by a
    considerable amount; high-$p_T$ data is hence clearly dominated by
    processes other than BE correlations.}
\end{figure*}

\begin{figure*}[t]
  \centerline{\epsfig{figure=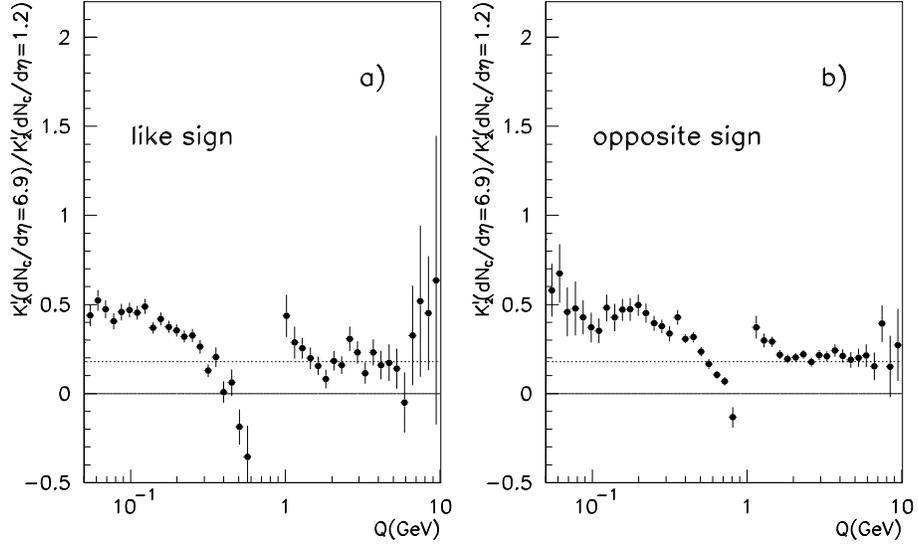,width=128mm}}
  \caption{
    The ratio of two $K_2^I(Q)$ corresponding to two selections of
    $dN_c/d\eta$ as indicated, for the low-$p_T$ sample (ii). The dashed
    line indicates the value of the ratio for the case that $K_2^I(Q)
    \propto 1/N_c$}
\end{figure*}

\begin{figure*} 
  \centerline{\epsfig{figure=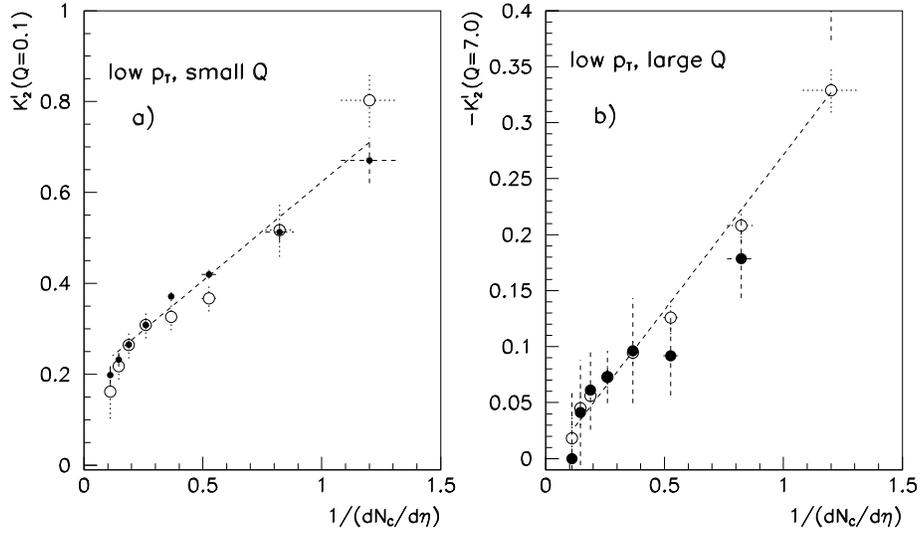,width=128mm}}
  \caption{
    a) Multiplicity dependence of $K_2^{I\ls}$ (filled circles) and
    $K_2^{I\os}$ (open circles), both at $Q=0.1$ GeV, b) as in a) but
    for the large $Q=7$~GeV-region.  Note from Fig.\ 1 that the
    cumulants are negative at large $Q$.  For better comparison of the
    respective dependencies on $dN_c/d\eta$, the absolute values have
    been scaled by constant factors. Dashed lines are best fits using
    Eq.\ (\ref{resin}).  }
\end{figure*}

\end{document}